\documentclass[aps,prl,reprint,amsmath,amssymb,amsfonts,floatfix,showpacs]{revtex4-1}
\usepackage{graphicx}
\usepackage{color}
\usepackage[caption=false]{subfig}
\begin{document}

\title{Sufficient minimal model for DNA denaturation: Integration of harmonic scalar elasticity and bond energies}
\author{Amit Raj Singh$^1$, Rony Granek$^{1,2}$}
\email{rgranek@bgu.ac.il}
\affiliation{$^1$ The Stella and Avram Goren-Goldstein Department of Biotechnology Engineering, and\\
$^2$ The Ilse Katz Institute for Meso and Nanoscale Science and Technology, Ben-Gurion University of The Negev, Beer Sheva 84105, Israel.}
\begin{abstract}

We study DNA denaturation by integrating elasticity -- as described by the Gaussian network model -- with bond binding energies, distinguishing between different base-pair and stacking energies. We use exact calculation, within the model, of the Helmholtz free-energy of any partial denaturation state, which implies that the entropy of all formed bubbles (``loops") is accounted for. Considering base-pair bond removal single events, the bond designated for opening is chosen by minimizing the free-energy difference for the process, over all remaining base-pair bonds. Despite of its great simplicity, for several known DNA sequences our results are in accord with available theoretical and experimental studies. Moreover, we report free-energy profiles along the denaturation pathway, which allow to detect stable or meta-stable partial denaturation states, composed of ``bubbles", as local free-energy minima separated by barriers. Our approach allows to study very long DNA strands with commonly available computational power, as we demonstrate for a few random sequences in the range 200-800 base-pairs. For the latter we also elucidate the self-averaging property of the system. Implications for the well known breathing dynamics of DNA are elucidated.

\end{abstract}

\pacs{87.15.-v, 87.14gk, 87.15A-, 87.10.Ca}

\maketitle
DNA carries the genetic information which is responsible for the development and functioning of all
living organisms, and its transcription and replication involve the opening and closing of the
double helix \cite{Alberts}. DNA is a highly dynamic molecule in which base pairs (bp's) fluctuate widely \cite{Wildes}. Besides the common thermal vibrations, bonded bp's can also locally disconnect and reconnect, a phenomenon termed ``breathing" motion. Breathing fluctuations are an example of fluctuations in a quasi one-dimensional system \cite{Fye} . They have been argued to control DNA protein binding, chemical reactivity, mutagenesis and cancerogenesis \cite{Frank} . At a certain temperature, or at a high concentration of denaturating solvent such as formaldehyde, this local separation of the two strands extends over the entire molecule resulting in a complete separation of the two strands, a phenomenon known as denaturation or melting \cite{Wartell}.

Thermally induced DNA denaturation has been investigated by using several experimental techniques,
including spectroscopic methods\cite{Cantor}, fluorescent energy transfer(FRET),
and circular dichroism(CD)\cite{Gelfand}, calorimetry\cite{Senior,Vesnaver,Chan},
and electrophoretic mobility assays\cite{Myers,Zhu}, and theoretical models, which consist of
phenomenological thermodynamic approaches and statistical mechanics models at various levels of approximation and coarse graining \cite{Newell,Hill,Zimn,Poland,Fisher,Poland1,Fixman,Azbel,Wada,Gotoh,Breslauer,Peyrard,Dauxois1,Sugimoto,Owczarzy,Campa,Chalikian,Dauxois,Kafri,Causo,Carlon,Cule,Singh,Santa,Blake1}. It has been shown that homosequence DNA denatures quite abruptly within a very small temperature interval,
whereas heterosequence DNA denatures {\it via} a few partial denaturation states \cite{Gotoh,Blake}, leading to a denaturation
curve (plot of average number of open bp's against temperature) whose shape is highly sensitive to the
sequence. The melting and its transition do not only depend on the fraction of strong G-C or A-T bonds, but
are also sequence specific \cite{Leijon}. This phenomena has been recently suggested as an application for DNA sequencing \cite{Chen}.
However, the melting curve alone does not allow to pin-point the presence or absence of intermediate states.
In addition, it does not provide information on the free-energy profile along the denaturation pathway at
different temperatures. This is necessary for studying fluctuations between different partial denaturation
states, whose kinetics is, in essence, the (above mentioned) DNA breathing motion. Such transitions can be studied by modern experimental techniques, e.g., fluorescence resonance energy transfer \cite{Gelfand} and fluorescence correlation spectroscopy \cite{Oleg}.

In this Letter we develop an extremely computationally cheap method that allows to study DNA denaturation
for long sequences and obtain the Helmholtz free-energy, and consequently the Boltzmann occupation
probabilities, along the melting pathway at different temperatures. We use an approach that was recently
developed to study protein unfolding \cite{Srivastava,Granek} and generalize it for DNA denaturation. The approach integrates
elasticity, as described by the Gaussian network model (GNM), with bond binding energies. In our
generalization for DNA, we assign different binding energies for different bp's (two energy parameters),
and different stacking (binding) energies for different pairs of bp's (sixteen energy parameters),
leading to total of eighteen energy parameters. Likewise, the GNM part is generalized to account for
the corresponding eighteen spring constants whose values are taken proportional to the binding energies,
and includes also the covalent (single strand ``backbone" bonds) spring constant.

dsDNA building blocks are single DNA strands, see Fig. 1. Each strand consists of a string of nucleotides,
each of which is mapped onto three interaction nodes: two backbone nodes, (i) one node for phosphate group (full circle), (ii) another node for sugar group (open circle), and (iii) a third node for the base (square), either adenine (A),
thymine(T), cytosine(C), or guanine(G). A denaturation state of DNA is determined by the bonds that have
been broken (or those that remain). At any given state, the elasticity is presented here by an elastic
network of harmonic springs, the GNM, in which the spring constants are assigned as follows.
Covalent bonds connect all backbone nodes, and the base nodes to their corresponding
backbone nodes; all are assigned in this simplified model an identical spring constant $\gamma m$, where $m$ is
dimensionless and $\gamma$ a reference spring constant. Next consider the (non-covalent) bonds
connecting nucleotides, representing nucleotide-nucleotide interactions, complementarity and stacking interactions. Complementarity interactions, associated with the formation of hydrogen bonds, induce the Watson-Crick pair
formation, i.e. A-T and G-C. The A-T base pair is formed by two hydrogen bonds and is assigned a
spring constant $\gamma a$ whereas G-C base pair is formed by three hydrogen bonds and assigned spring constant $\gamma b$. Stacking interactions are between pairs of bp's and are responsible for their preference to lie on top of each other, spring constant $\gamma d_{\nu}$, where $d_{\nu}$
varies between the different, $\nu=1,...,16$, possible stacking arrangements of pairs of bp's.
The values of $a$, $b$, and $\{d_{\nu}\}$, representing the relative spring constants, are assumed
proportional to the binding energy $\epsilon$ of the corresponding bonds (see below). This corresponds to the basic assumption that there is a uniform bond cutoff (stretching) distance, $x_c$, for all breakable bonds in the system such that $\epsilon=(1/2)\gamma x_c^2$ for each bond. This leads to the relative values presented in Table S3 in the SI. Since the stacking interactions are presented by two springs connecting two base-pairs (see Fig. 1), each such spring is assigned half of the value of the total ``spring" associated with this interaction (such that, e.g., $a/d_{\nu}=2(\epsilon_{AT}/\epsilon_{\nu})$ and $b/d_{\nu}=2(\epsilon_{GC}/\epsilon_{\nu})$). Note that we conveniently assigned for the weakest spring $d_{16}=1$, however, this choice has no affect on any of our results so long as all ratios of spring constants (i.e. values such as $a/d_{16}$, $b/d_{16}$ etc.) remain the same. This is true because we can always multiply $\gamma$ by $d_{16}$ while dividing $\{\Gamma_{ij}\}$ by $d_{16}$, see Eq. (1), and, as argued below, the value of $\gamma$ does not affect denaturation properties. The value of $m=10$ was chosen as the typical ratio of covalent to hydrogen bond spring constants.

\begin{figure}[t]
	\begin{center}
		\includegraphics[width=0.27\textwidth]{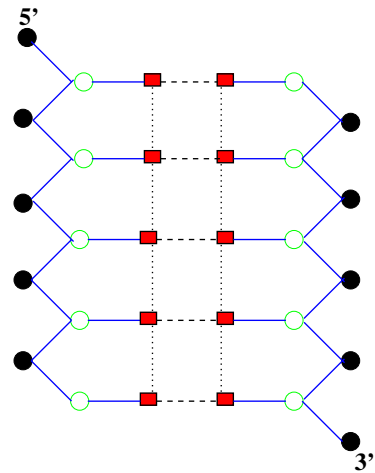}
		\caption{(Color online) Schematic representation of the model. Full (black) circles, open (green) circles and (red) squares represent
the phosphate, sugar and base, respectively. Solid (blue), dashed, and dotted lines represent the covalent bonds, hydrogen bonds and
stacking interactions, respectively. The elasticity of the stacking interactions (dotted lines) is divided into two parallel springs.}
		\label{model}
	\end{center}
\end{figure}
\begin{table}[tbp]
\centering
\caption{Base-pair stacking energies for sixteen possible DNA base-pair stacking patterns based on quantum
chemistry calculations~\cite{Ornstein} in units of kcal/mol. Each single value corresponds to a pair of dotted lines in Fig. 1.}
\label{my-label}
\begin{tabular}{|c|c|c|c|c|c}
\hline
\begin{tabular}[c]{@{}c@{}}DNA Dimer\end{tabular} & Stacking energy & \begin{tabular}[c]{@{}c@{}} DNA Dimer \end{tabular} & Stacking energy \\
\hline
\begin{tabular}[c]{@{}c@{}}  C$\cdot$G \\ G$\cdot$C \end{tabular}               & $\epsilon_1$ = 14.59                              & \begin{tabular}[c]{@{}c@{}} C$\cdot$G \hspace{.3cm}  T$\cdot$A\\  A$\cdot$T \hspace{.3cm}  G$\cdot$C\end{tabular} & $\epsilon_2$,
$\epsilon_3$ = 10.51 \\
\hline
\begin{tabular}[c]{@{}c@{}}C$\cdot$G \hspace{.3cm}   A$\cdot$T\\ T$\cdot$A \hspace{.3cm}   C$\cdot$G\end{tabular}      & $\epsilon_4, \epsilon_5$ = 9.81                              & \begin{tabular}[c]{@{}c@{}}G$\cdot$C\\ C$\cdot$G\end{tabular}                  & $\epsilon_6$ = 9.69                              \\
\hline
\begin{tabular}[c]{@{}c@{}}G$\cdot$C \hspace{.3cm}  C$\cdot$G\\ G$\cdot$C \hspace{.3cm}    C$\cdot$G\end{tabular}    & $\epsilon_7,\epsilon_8 $ = 8.26                              & \begin{tabular}[c]{@{}c@{}}T$\cdot$A\\ A$\cdot$T\end{tabular}                  & $\epsilon_9 $ = 6.57                              \\
\hline
\begin{tabular}[c]{@{}c@{}}G$\cdot$C \hspace{.3cm}   A$\cdot$T\\ T$\cdot$A \hspace{.3cm}   C$\cdot$G\end{tabular} & $\epsilon_{10}, \epsilon_{11} $ = 6.57                              & \begin{tabular}[c]{@{}c@{}}G$\cdot$C \hspace{.3cm}    T$\cdot$A\\ A$\cdot$T \hspace{.3cm}   C$\cdot$G\end{tabular}     & $\epsilon_{12}, \epsilon_{13} $ = 6.78                              \\
\hline
\begin{tabular}[c]{@{}c@{}}A$\cdot$T \hspace{.3cm}   T$\cdot$A\\ A$\cdot$T \hspace{.3cm} T$\cdot$A\end{tabular}  & $\epsilon_{14}, \epsilon_{15} $ = 5.37                               & \begin{tabular}[c]{@{}c@{}}A$\cdot$T\\ T$\cdot$A\end{tabular}                  & $\epsilon_{16} $ = 3.82                                \\
\hline
\end{tabular}
\end{table}

We denoted by $L$ the strand length, i.e. no. of bp's. The GNM Hamiltonian of a DNA having $N=6L$ nodes, either in the native state or at any intermediate denaturation state, can be written as
\begin{eqnarray}
 H_{GNM} =\frac{1}{2}\gamma \sum_{<ij>}\Gamma_{ij}(\vec{u_i}-\vec{u_j})^2
\label{equ1}
\end{eqnarray}
where $i$ is node index, and the sum runs over all node pairs $<ij>$. Here $\vec{u_i}$ are the displacement vectors in three-dimensional (3D) space, and $\{\Gamma_{ij}\}$ are elements of the connectivity (spring constant) matrix, and take the values
\begin{equation}
\Gamma_{ij}=
\begin{cases}
 m \hspace{.3cm}{\text {for}} \hspace{0.2cm}  \hbox{covalent-bond}  \\
 a \hspace{.45cm}{\text {for}} \hspace{0.2cm}  \hbox{A-T base pair} \\
 b \hspace{.45cm} {\text {for}} \hspace{0.2cm} \hbox{G-C base pair} \\
 d_{\nu} $\hspace{.25cm}$ {\text {for}} \hspace{0.2cm} \hbox{stacking-interaction} \\
\end{cases}\;,
\label{eq2}
\end{equation}

To describe the free-energy difference between each denaturation state we need also to assign binding energies to bonds. Covalent bond rupture is not relevant here thus no binding energy is assigned to it. Base-pair binding energies are presented by -$\epsilon_{AT}$ and -$\epsilon_{GC}$ for A-T and G-C, respectively, with
$\epsilon_{AT}, \epsilon_{GC} >0$, and stacking binding energies by $-\epsilon_{\nu}$, $\nu=1,...,16$. Their values have been calculated using quantum chemistry methods \cite{Ornstein,Danilov}. In our computational study we used $\epsilon_{AT}=7$ kcal/mol, $\epsilon_{GC}=16.79$ kcal/mol, and the values given in Table I for the stacking energies. We define for convenience the zero of energy scale to be at the native state.

Base-pair bonds will be allowed to break independently, however stacking bonds are not independent and only follow the base-pair bonds. We shall assume that stacking bond breakage occurs when at least one participating base-pair is broken \cite{foot-stacking}. Thus, two stacking bonds are broken when an internal base-pair breaks, whereas only one stacking bond is lost when an end base-pair breaks. Therefore, assigning a ``Lattice-Gas" (``Ising") type variable to a base-pair {\it bond}, $S_{ij}=0,1$ for a broken or formed bond, respectively, the Hamiltonian of the system at any partial denaturation state, where $n_1$ A-T bonds and $n_2$ G-C bonds have been {\it already broken}, is
\begin{eqnarray}\nonumber
H(n_1,n_2, \{S_{ij}\})&=& H_{GNM}+n_1\epsilon_{AT}+n_2\epsilon_{GC}\\
&& +\sum_{<ijk\ell>}\epsilon_{ijk\ell} \left(1-S_{ij}S_{k\ell}\right)
\end{eqnarray}
where $\epsilon_{ijk\ell}\equiv \epsilon_{\nu}$ represents the corresponding stacking energy of the specific pair of bp's defined in Table I, and the sum runs over (nearest neighbor) base quartets. Note that $n=n_1+n_2=\sum_{<ij>,\text{bp}} 1-S_{ij}$ is the total number of broken base-pair bonds, $0\leq n\leq L$.

For compactness we now transform Eq. (1) into a standard quadratic form
\begin{equation}
H = \frac{1}{2}\gamma\sum_{m = x,y,z} {\bf u_m^T  \cdot \Lambda \cdot u_m},
\end{equation}
where ${\bf u_m}$'s are vectors of length $N$ in node space. The elements of the matrix $\bf{\Lambda}$ are

\begin{equation}
\Lambda_{ij}=
\begin{cases}
 -\Gamma_{ij} \hspace{1.4cm} {\text {if}} \hspace{0.3cm} i\neq j \\
\displaystyle\sum_{k\ne i} \Gamma_{ik} \hspace{1cm} {\text {if}} \hspace{0.3cm}i=j
\end{cases}\;.
\end{equation}

In order to compute the Helmholtz free-energy $F$ corresponding to the Hamiltonian Eq. (4), we first rewrite the GNM Hamiltonian associated with one of the axes $m$ $(m = x,y,z)$ in eigenmode space
\begin{equation}
H_{GNM}^{(m)} = \frac{1}{2}\gamma \sum_{\alpha} \lambda_{\alpha}A_{\alpha}^2,
\end{equation}
where $\lambda_{\alpha}$ is the eigenvalue corresponding to mode $\alpha$ of the  matrix ${\bf \Lambda}$, and $A_{\alpha}$ is the amplitude of this mode. Calculating the canonical partition function $Z$ of a given intermediate state is a textbook exercise and is easily performed since it involves only Gaussian integrals \cite{Schulmann}.
This allows to obtain the Helmholtz free-energy of an intermediate denaturation state, $F=-k_BT \ln Z$, as the sum of bond binding energy and an elastic (GNM) free-energy,
\begin{eqnarray}\nonumber
F&=& n_1\epsilon_{AT}+n_2\epsilon_{GC}+\sum_{<ijk\ell>} \epsilon_{ijk\ell} \left(1-S_{ij}S_{k\ell}\right)\\
&& + {3\over 2}k_BT\sum_{\alpha}\ln(\lambda_{\alpha}) - {3\over 2} (N-1) k_BT\ln\left(\frac{2\pi k_BT}{\gamma\lambda_B^2}\right) .
\label{Ftot}
\end{eqnarray}
where $N=6L$ is the total number of nodes in the system, and $\lambda_B=(h^2/(2\pi mk_BT))^{1/2}$ is the De Broglie wavelength. Note that the last term, which involves $\gamma$, is independent of DNA base-pairing configuration and therefore cannot affect the denaturation sequence. It should be included only when calculating measurable thermodynamic properties such as specific heat, where it adds a temperature dependent term which is however independent of DNA base-pairing configuration.

The fourth term is essentially the relative configurational entropy contribution to the free-energy (i.e. $-TS$), and
is specific to the network topology. As, base-pair bonds, and their associated stacking bonds, break, eigenvalues decrease, implying increase of entropy, which drives the denaturation.
Indeed, as local ``bubbles'' appear, each of the two strands can move freely within a bubble (subject to end constraints),
implying an increase in entropy. We have omitted in Eq. (7) a temperature dependent constant
(see Eq. (A11) in Ref. 43) since it is independent of network topology and so cannot affect the pathway of denaturation.

To obtain the thermodynamic equilibrium configuration of the DNA at a given temperature $T$,
the free-energy Eq. (7) has to be minimized over all possible $2^{L}$ network topologies,
namely over all network topologies at any given $n$, $n=0,...,L$. Since this becomes a computationally formidable
task for $L>>1$ \cite{foot-note}, we adopt a steepest-descent-like procedure that can be described as follows:
(i) assume a particular (partially denatured) DNA and
calculate the free-energy difference associated with a single base-pair break, for each of the remaining
bp's in the network, (ii) choose the base-pair for which this free-energy difference is minimal, and break it
together with the stacking bonds associated with this base-pair, and  (iii) repeat steps (i)-(ii) for $n=0,...,L$.
Using this procedure we consider only $\frac{L(L+1)}{2}-1$ partial denaturation states describing the most
probable denaturation pathway \cite{foot-note}.
Once the opening sequence of bp is found, it is possible to use Eq. (7) to plot the free-energy ``landscape"
along this pathway at various temperatures. At temperatures much below the melting temperature/s -- while
free-energy will increase with denaturation -- this pathway represents the most probable denaturation pathway
that can be achieved by thermal fluctuations. At temperatures much above the melting temperature/s, free-energy will descend along the pathway,
and denaturation will occur as an instability (on the very same pathway). At intermediate temperatures, a rich
free-energy profile may exist that may reveal partial denaturation states.
From Eq. (7), we can compute the free-energy difference, at constant temperature $T$, between a network with $n+1$ broken bp's
and an $n$ broken bp network, $\delta F(n)=F(n+1)-F(n)$. We obtain
\begin{equation}
\delta F(n)  = \epsilon^{**} - \frac{3}{2}k_BT \left(\sum_{\alpha, old}\ln\lambda^{(n)}_{\alpha} - \sum_{\alpha, new}\ln\lambda^{(n+1)}_{\alpha}\right),
\label{deltaF}\end{equation}
\begin{figure}[h]
\subfloat[]{%
  \includegraphics[clip,width=0.7\columnwidth]{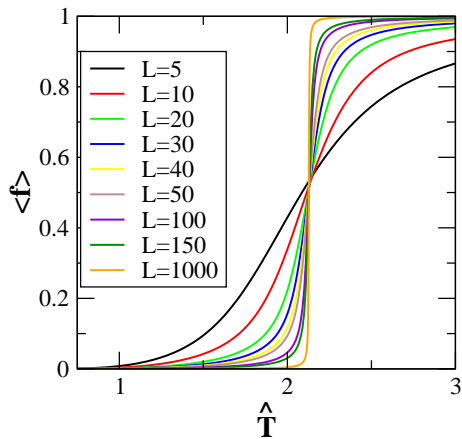}%
}

\subfloat[]{%
  \includegraphics[clip,width=0.7\columnwidth]{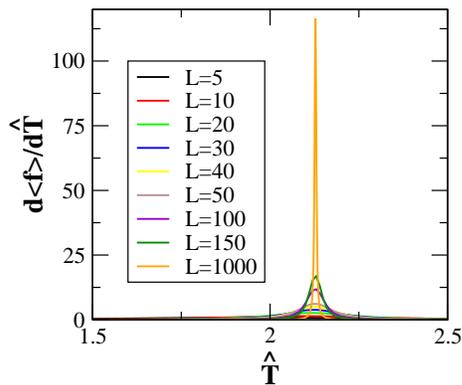}%
}

\caption{(a) Melting profile for dGdC DNA for a range of strand lengths $L$ between 5 and 1000. We can observe that the melting curve sharpens with increasing $L$. (b) Derivative of melting profile. The peak coincides with the melting temperature and sharpens with increase of $L$. These trends are generally expected for first order phase transitions in finite size systems, and similar to the findings of Ref. [49]. The variation of melting temperature $\hat T_m$ at different $L$'s is shown in Fig. S7 (in the SI).}

\end{figure}

where $\epsilon^{**}$ is the energy difference between the (specific) $n+1$ unbonded base-pair network and
the (specific) $n$ unbonded base-pair network. Thus $\delta F(n)=0$ defines the temperature for which associated
bond becomes unstable against breakage, which will allow us to construct an instability phase diagram.
\begin{figure}[h]
        \begin{center}
                \includegraphics[width=.40\textwidth]{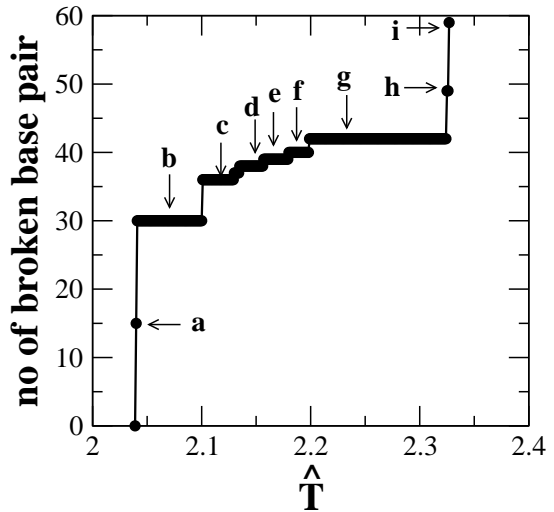}
                \caption{Denaturation instability phase diagram for L60B36 along the denaturation pathway:
Number of broken base-pair bonds against reduced temperature $\hat T=T/T_0$. Different denaturation states of
L60B36 are denoted by $a-h$. We observe several jumps where in each jump several base-pair bonds break simultaneously due to an avalanche-like effect.}
                \label{instability}
        \end{center}
\end{figure}

We study a large variety of DNA sequences that have received much theoretical attention in the past decade,
which is possible due to the great efficiency of our method, see the supporting information
(SI) for a detailed description. Our results are in complete accord with available theoretical and experimental results \cite{Inman,Marmur} . Moreover, we included in these studies long chains upto 1000 bp's, that have not been studied thus far. Importantly, for homo-sequences we observe that as the sequence length increases, the transition becomes sharper. This is exemplified by the melting (or denaturation) curve, see, e.g., Fig. 2. This is the expected trend, as true phase transition can exist only for infinite systems, whereas for finite systems finite size effects exist. Likewise, for diblock sequences made of A-T and G-C blocks, there are two transitions whose sharpness increases with increasing length of each of the blocks (Fig. S10 in the SI).

Here we would first like to focus on one particular sequence of special interest. It is a heterogeneous DNA sequence, i.e. a sequence that contains both G-C and A-T bp's, $\bf L60B36$ \cite{Zeng,Peyrar}.
The sequence of this strand is CCGCCAGCGGCGTTATTACATTTAATTCTTAAGTATTATAAGTAATATGGCCGCTGCGCC, and
we shall index the bp's from 1 to 60 from left to right; the $(5')$-end is on the left and the $(3')$-end
is on the right. The specific sequence of denaturation pathway that we obtain is presented in the SI (File S111).
For ease of presentation we define a reference temperature $T_0$ such that $k_BT_0=\epsilon_{16}$ (see Table I).
Fig. 3 shows the instability phase diagram (see discussion after Eq. (8)), showing the reduced temperature
$\hat T=T/T_0$ of instability breakup of base-pair bonds, along the denaturation pathway.
Note that denaturation is neither continuous, nor abrupt. Rather, it occurs in a few jumps between intermediate denaturation states.

\begin{figure}[ht]
        \begin{center}
                \includegraphics[width=.40\textwidth]{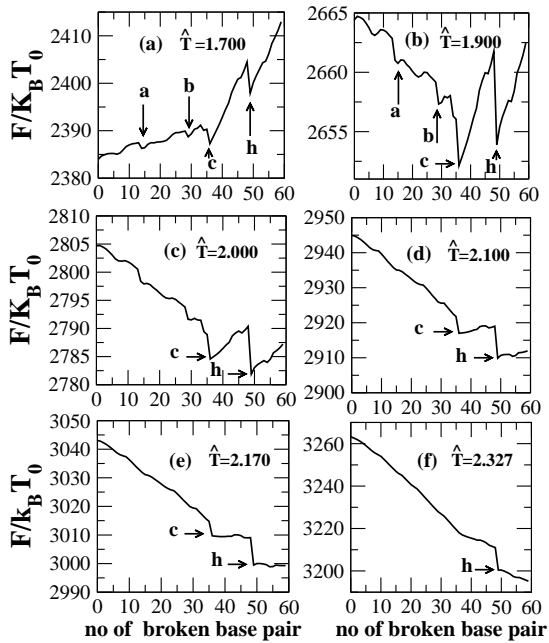}
                \caption{Free-energy profile for L60B36. The dimensionless free energy is plotted against
the number of open bp's at different temperatures.}
                \label{free-energy}
        \end{center}
\end{figure}
In Fig. 4 we present the free-energy profile (i.e. 1D free-energy landscape) along the denaturation pathway
at different (reduced) temperatures. The reduced free-energy, $F/k_B T_0$, is plotted against the number of
open base-pair bonds along the denaturation pathway. At $\hat T\simeq 1.700$, Fig. 4(a), the global
minimum appears at the native state, implying that it is the most thermodynamically stable state.
We also observe a few local minima separated by energy barriers, corresponding to meta-stable,
partial denaturation, states. Some of these states, i.e. {\it a, b, c} and {\it h}, correspond
to states observed in the instability phase diagram, Fig. 3. As the temperature is raised, states change
role between metastability and stability -- e.g., state {\it c} is the most stable state at $\hat T=1.9$
(Fig. 4(b)) and state {\it h} is most stable at $\hat T=2.0$ (Fig. 4(d)) -- until at elevated
temperatures the complete denaturation state becomes the most stable one.

In Fig. 5 we demonstrate schematically the DNA configuration in the dominant partial denaturation
states {\it a,b,c} and {\it h}.
\begin{figure}[ht]
        \begin{center}
                \includegraphics[width=.40\textwidth]{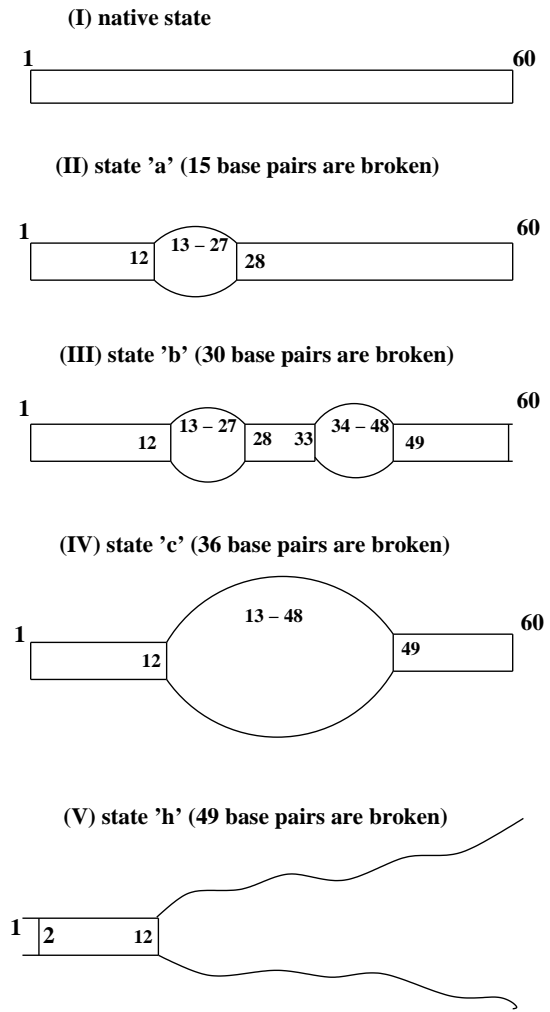}
                \caption{Schematic representation of
partial denaturation states of L60B36 (`native', `a', `b' `c' and `h') marked by arrows in Fig. 3. (II)
Upto state {\it a}, total of $15$ native base-pair contacts (from $13^{th}$ to $27^{th}$) are removed, (III) from
state {\it a} to {\it b}, additional $15$ native base-pairs (from $34^{th}$ to $48^{th}$) disconnect, (IV) from {\it b} to {\it c}, more $6$
native base-pair bonds (from $28^{th}$ to $33^{th}$) break, and
(V) from {\it c} to {\it h}, additional $13$ native base-pairs disconnect ($1^{st}$ base pair and from $49^{th}$ to $60^{th}$).}
                \label{state-diagram}
        \end{center}
\end{figure}
Physically, state {\it c} corresponds to broken $36$ bp's,
from the $13^{th}$ to $48^{th}$ (termed previously as a ``bubble at middle",\cite{Zeng,Peyrar} ),
Fig. 5.IV. Fig. 5.V schematically depicts state {\it h} showing the unzipping (from the right)
of bp's 13-60. We note that while state {\it c} has been theoretically predicted \cite{Peyrar} (using other methods)
and experimentally discovered \cite{Zeng} , other states were not mentioned. In particular, the existence
state {\it h}, which is the most stable state in a quite wide temperature range ($\hat T\simeq 1.940-2.140$ ),
was not predicted thus far. In Fig. 6 we complement the free-energy plots by plotting the Boltzmann
state occupation probability. We observe peaks corresponding to local minima in Fig. 4, as expected.
Since at various temperatures we find free-energy levels of meta-stable states that are close to the global minimum,
their population is found non-negligible.
\begin{figure}[h]
        \begin{center}
                \includegraphics[width=.40\textwidth]{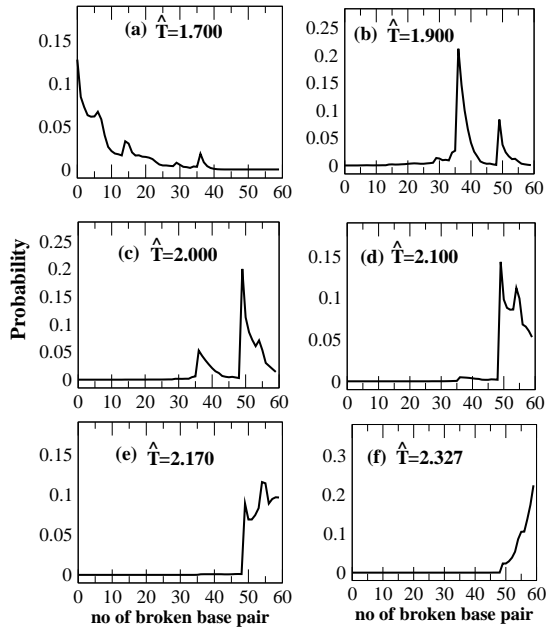}
                \caption{Boltzmann occupation probability for  L60B36 plotted against the number of opened
base pairs at different temperatures.}
                \label{fig:example}
        \end{center}
\end{figure}

\begin{figure}[h]
        \begin{center}
        \includegraphics[width=.40\textwidth]{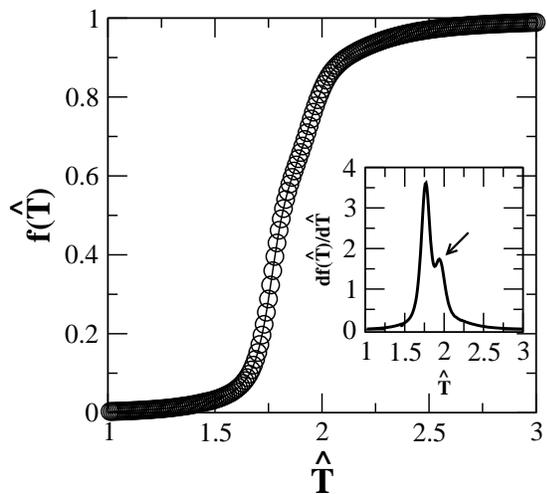}
                \caption{Melting profile of the sequence $L60B36$. The mean number of broken base-pairs $f(\hat T)$ is plotted against reduced temperature $\hat T$. The inset shows the derivative of the curve. }
                \label{meltingprofile}
        \end{center}
\end{figure}

To end this analysis, we show in Fig. 7 the melting curve (calculated using
the Boltzmann probabilities) and its derivative. Note the double shoulder/double peak in the curve/curve derivative.
Surprisingly, this behavior was not found in previous studies \cite{Peyrar}, where only one shoulder (one peak)
was observed, despite of the prediction of state {\it c}. We suggest that the presence of state {\it h},
in addition to the high stability of state {\it c}, is the cause of this behavior. In the SI (Sec. I) we report results for another heterosequence, where a double shoulder is also observed in the denaturation curve, however, it is caused by a single dominant partial denaturation state, a ``bubble at the end" \cite{Zeng, Peyrar, Montri}.

\begin{figure}[h]
        \begin{center}
        \includegraphics[width=.50\textwidth]{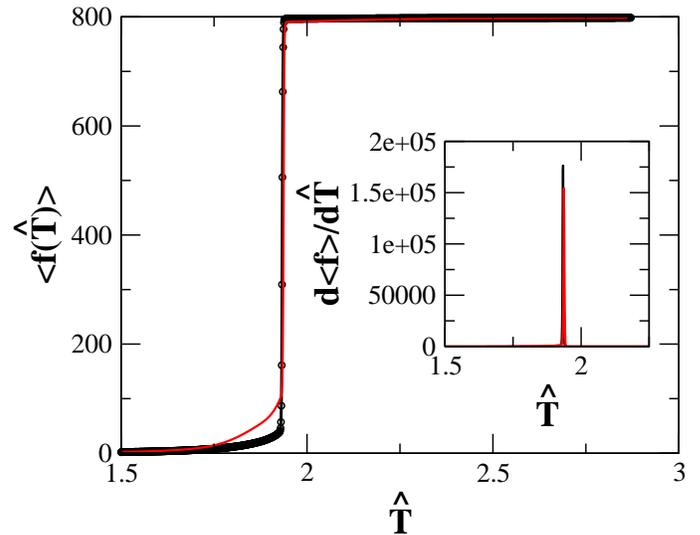}
                \caption{Melting profile for two realizations of a long random sequence (800 base pairs) with 50\% A-T and 50\% G-C. The inset shows the derivative of the melting curves. Black and red lines correspond to the base-pair sequences detailed in files S111 and S222 in the SI, respectively.}
                \label{meltingprofile}
        \end{center}
\end{figure}

\begin{figure}[h]
        \begin{center}
        \includegraphics[width=.50\textwidth]{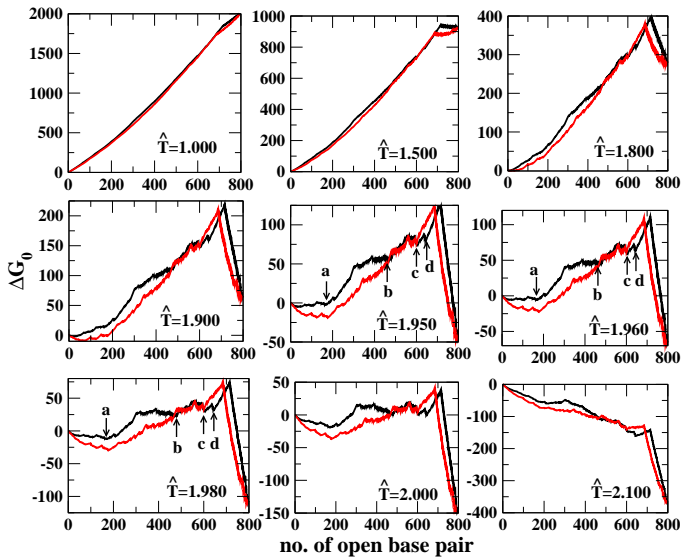}
                \caption{Free-energy profiles for the two realizations of a long random sequence (800 base pairs) described in the captions of Fig. 8 (corresponding colors). The dimensionless free-energy difference between partially denaturation and native states is plotted against the the number of open bp's at different reduced temperatures. Arrows indicate local minima at n=171, n=462, n=602 and n=642 for one of the realizations (black line, file S111 in the SI). }
                \label{meltingprofile}
        \end{center}
\end{figure}

To show that our model is suitable for studying long sequences, we have also implemented it on a few long random sequences. First, we considered two realizations of a long random sequence (800 base pairs), with 25\% chance for each base in a single stand, resulting in 50\% chance for A-T base pair and 50\% chance for G-C pair; details of the specific sequences studied are shown in the SI (Files S222, S333). We note in passing that we may expect that the melting curve of long random sequences is ``self-averaging"\cite{Orlandini} , such that no difference is expected in the melting profile of the ``infinite system" when changing from one realization of the sequence to another. Figs. 8 and 9 present the melting curve and free-energy profile along the denaturation pathway at different (reduced) temperatures for the two sequences (realizations of the random sequence). Indeed, we see that the two melting curves are almost identical, and yield indistinguishable melting temperatures. The slight difference between the curves prior to the transition may be attributed to finite size effects. This overlap of the melting curve occurs despite the difference in free-energies shown in Fig. 9. Figs. 10 and 11 show, for one of the realizations (black melting curve and free-energies in Figs. 8 and 9) the state of base pairs (open/close) in different partial intermediate states which are marked in Fig. 9. The red circles on top show the open states of base pair whereas the red circles on bottom show the close states of base pair.

Second, we analyzed two long random sequences, with different G-C concentration, L516(44\% G-C) and L234(59\% G-C). Deterministic sequences of the same length and G-C concentration have been previously studied experimentally\cite{Wartell} , with information about the actual sequence missing. Yet, it is still of interest to study the equivalent random sequences. In Fig. 12 we show the melting curves and their derivatives for a single realization of each of the sequences. Note that higher G-C concentration leads to a higher melting temperature, as found for the deterministic sequences studied above and in the SI. However, here again we only find a single melting temperature, just as in the 800 bp 50\% G-C sequence studied above. In the SI, Fig. S11  we present the resulting melting curves from MELTISM\cite{software} for exactly the same realization we randomly drew, showing very similar results.


As mentioned, in all our comparisons with known experimental and state-of-the-art theoretical studies, we find the same denaturation pathways. It could be instructive to make also a quantitative comparison for the melting temperature. This is presented in the SI, Tables S1 and S2, where we compare, for different sequences, ratios of melting temperatures of different pairs of sequences calculated using our model to those obtained from the MELTISM software \cite{software} and experiment\cite{Zeng, Alexandrov}. We find, in general quite good agreement, suggesting that the approximations involved in our model do not affect much the accuracy.
\begin{figure}[t]
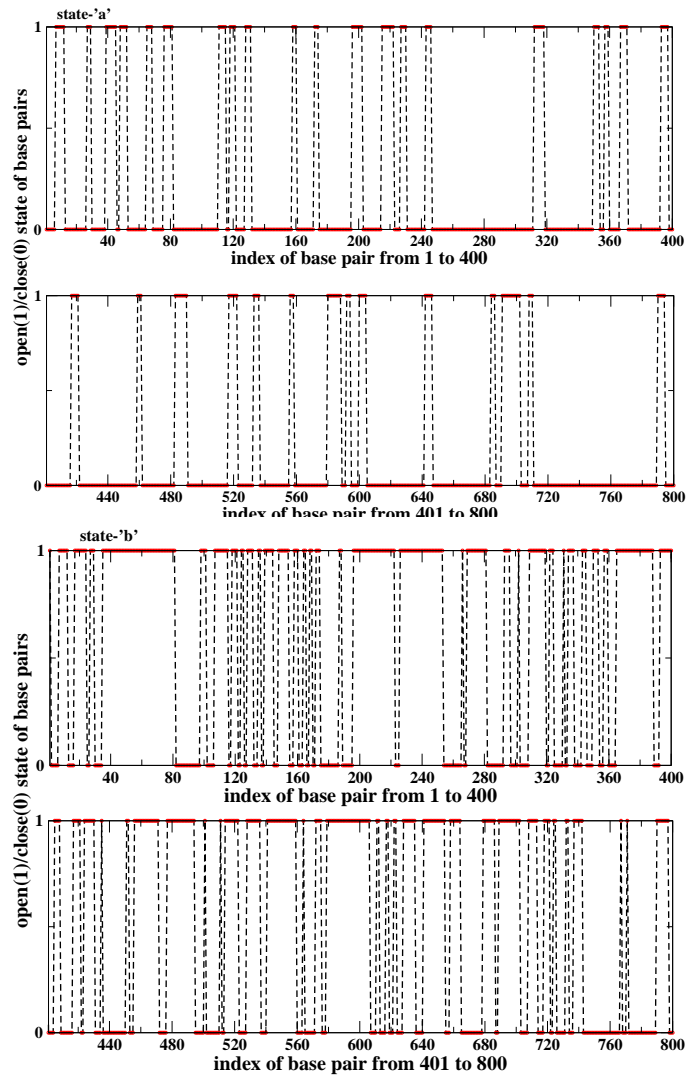

        \begin{center}
        \includegraphics[width=.50\textwidth]{jcp10a.eps}
        \includegraphics[width=.50\textwidth]{jcp10b.eps}
                \caption{Representation of
partial denaturation states (`a', `b') of long random sequence DNA marked by arrows in Fig. 9.
`a' and `b' state with local minima at n=172 and n=462 respectively.
Open and close state of base pair presentes by 1 and 0 respectively along the base pair index.
}
                \label{meltingprofile}
        \end{center}
\end{figure}

\begin{figure}[h]
        \begin{center}
        \includegraphics[width=.50\textwidth]{jcp11a.eps}
        \includegraphics[width=.50\textwidth]{jcp11b.eps}
                \caption{Representation of
partial denaturation states (`c', `d') of long random sequence DNA marked by arrows in Fig. 9.
`c' and `d' state with local minima at n=602 and n=642 respectively.
Open and close state of base pair presentes by 1 and 0 respectively along the base pair index. }
                \label{meltingprofile}
        \end{center}
\end{figure}
\begin{figure}[h]
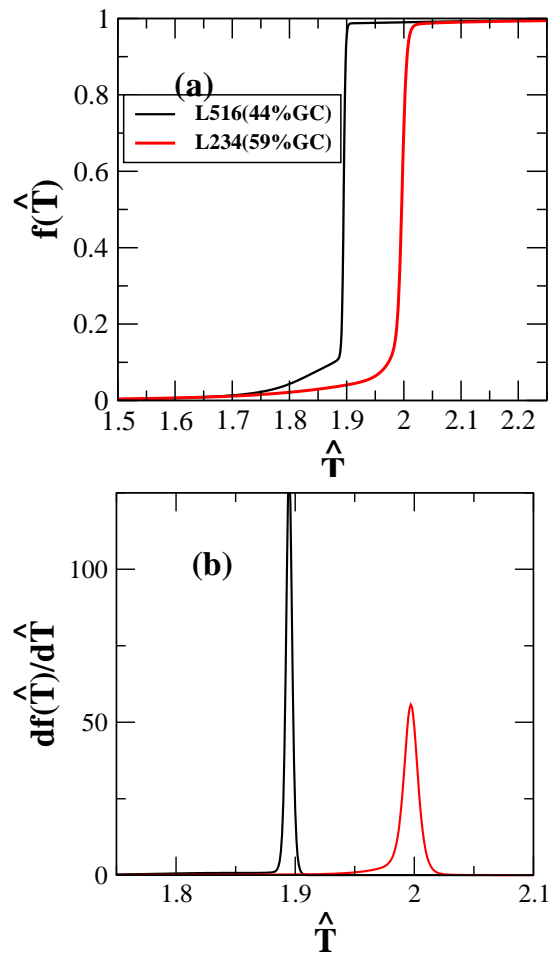

        \begin{center}
        \includegraphics[width=.40\textwidth]{jcp12a.eps}
        \includegraphics[width=.40\textwidth]{jcp12b.eps}
                \caption{ Fig. 12(a) shows melting profile for two random sequence of L234 and L516 basepair with 59\% G-C and 44\% G-C. Fig.12(b) shows the derivative of the melting curves. Red and black lines correspond to the L234 dsDNA and L516 dsDNA sequences detailed in the SI, respectively.}
                \label{meltingprofile}
        \end{center}
\end{figure}

To summarize, we have developed an extremely simplified off-lattice model for DNA denaturation that combines binding energies with elasticity. Here we reported on denaturation properties of two types of sequences. First, we studied special short sequences (L60B36 reported here, and L48AS21 reported in the SI) for which we found that even though the denaturation curves suggest a single intermediate state, our free-energy profiles {\it unambiguously show} that there are two such dominant states. Second, we considered two realizations of a long random sequence (800 bp) for which we demonstrated convergence of the two melting curves suggesting ``self-averaging" property of the system, as can be expected on general grounds. Evidently, since within the model the elastic canonical partition function of any partial denaturation state is calculated exactly, our theory includes the entropy of all formed bubbles (``loops"), which henceforth does not need direct evaluation, e.g., as in the Poland-Scheraga model\cite{Hill} and its variations. Despite of its simplicity, and the fact that our model focuses only on the most probable denaturation pathway (``steepest descent" type approach), the model predictions are in accord with most previous studies of popular sequences, several of them are reported in the SI. There are two main theoretical advantages to our method over many other methods. First, it is enormously computationally cheap, since an analytic general expression for the free-energy is implemented (see Eq.\ (\ref{Ftot}). This allows to study significantly long DNA strands (e.g., the 800 random base-pair strand studied here), which is important both practically and for the study of the infinite chain limit. Second, it allows to readily compute the free-energy along the sequence of denaturation pathway at various temperatures. The latter enables to clearly detect partial denaturation states as local minima separated by barriers, or as peaks of Boltzmann occupation probabilities. Further use of our approach can be for: (i) description of DNA breathing motion as kinetic transitions on the free-energy landscape (``random walk") between partial denaturation states, and (ii) Rouse and Zimm dynamics of DNA ensemble at various temperatures. These we defer for future studies.

We thank A. Srivastava and M. Feingold for useful discussions, and acknowledge partial financial support from the Deutsch-Israelische Projektkooperation (DIP), and from the Focal Technological Area Program of the Israeli National Nanotechnology Initiative (INNI).


\begin{thebibliography}{}
\bibitem{foot-stacking} We have checked that our assumption is justified by allowing stacking bonds to
break independently, which does not change our results.
\bibitem{foot-note} e.g., for $L \approx 20$ it takes about 6 cpu days to minimize the free-energy over the
whole $2^L$ states. Our approach reduces this time to $\sim$ 2 minutes.
\end{thebibliography}
\end{document}


\title{Electronic Supporting Information\\ `` Sufficient minimal model for DNA denaturation: Integration of harmonic scalar elasticity and bond energies''}
\author{Amit Raj Singh$^1$ and Rony Granek$^{1,2}$}
\affiliation{$^1$ The Stella and Avram Goren-Goldstein Department of
Biotechnology Engineering, Ben-Gurion University of The Negev,
Beer Sheva 84105, Israel.\\
$^2$ The Ilse Katz Institute for Meso and Nanoscale Science and Technology,
Ben-Gurion University of The Negev, Beer Sheva 84105, Israel.}
\maketitle

\section{I. Denaturation of ``L48AS21'' DNA}

We have chosen two known sequence  ``L48AS21'' and ``L60B36''. The results for L60B36 are reported in the main paper.
Here, we report results for L48AS21: $^{5'}$CATAATACTTTATATTTAATTGGCGGCGCACGGGACCCGTGCGCCGC$C^{3'}$.
The base pair index is assigned from 1 to 48 from left to right respectively. Fig. S1 shows the instability phase diagram along the denaturation pathway. Fig. S2 shows the free-energy profile at different temperature, where
the reduced free-energy is plotted against the number of open base pairs.
\begin{figure}[h]
        \begin{center}
                \includegraphics[width=.35\textwidth]{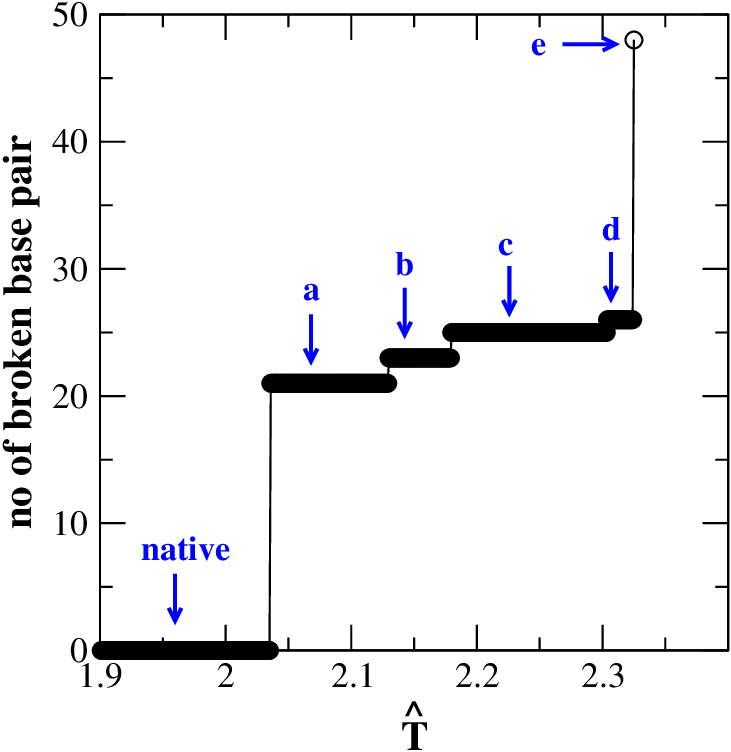}
                \caption{Fig. S1 shows denaturation instability phase diagram for L48AS21. Number of opened base pairs is plotted
against temperature. Different denaturation state of L48AS21 are denoted by {\it a-e}. We observe five jumps where in each
jump several base pair bonds break simultaneously due to an avalanchelike effect. (i) from native state to state $a$, total of $21$ native base pair contacts get opened, (ii) from state {\it a} to {\it b}, $2$ native base pair contacts,
(iii) from {\it b} to {\it c}, $2$ native base pair contacts, (iv) from {\it c} to {\it d}, $1$ native base pair contacts,
(v) from {\it d} to {\it e}, $12$ native base pair contacts get opened}
                \label{fig:example}
        \end{center}
\end{figure}

At $ \hat T\simeq 1.600$, Fig. S2(a), the global minimum appears at the native state, implying that it is the
most thermodynamically stable state. we also observe a few partial denaturation states separated by
energy barriers, corresponding to meta-stable states. As the temperature is raised states change role between metastability and stability -- e.g, state {\it a} becomes the most stable state in the temperature range
$1.65\lesssim \hat T\lesssim 2.15$ -- until at elevated temperatures the complete denaturation state becomes the most stable one.
At $\hat T\simeq 2.327$ the whole profile is monotonously decreasing from native state to fully
denaturation state (with no local minima in between), suggesting that the native state of the DNA is unstable against complete denaturation. Moreover, to complement the free-energy plots, we plot in Fig. S3 the
Boltzmann state occupation probability at the same temperatures shown in Fig. S2. We observe peaks corresponding to the local minima in Fig. S2, as expected.

\begin{figure}[ht]
\centering
\includegraphics[width=.40\textwidth]{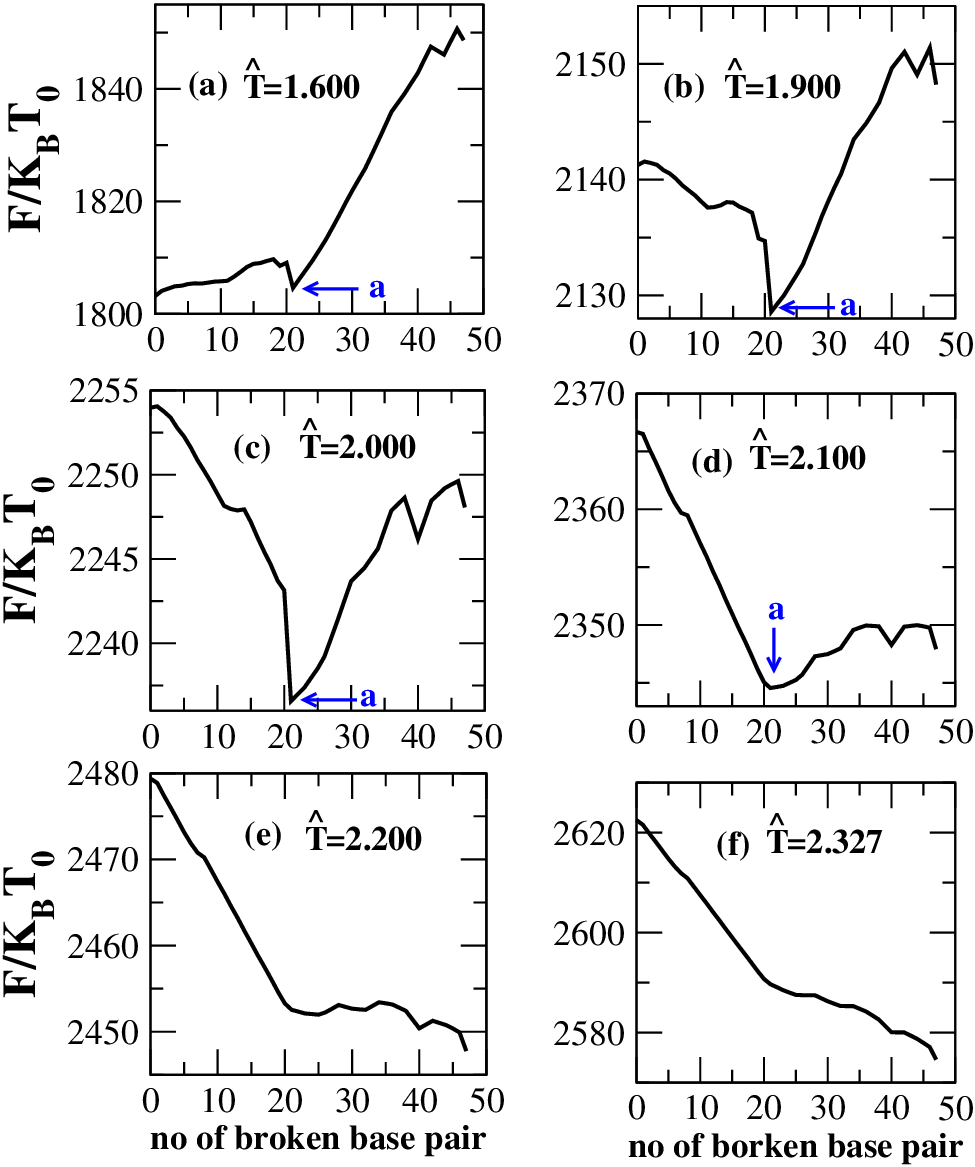}
\caption{Free-energy profile for L48AS21.The dimensionless free energy is plotted against
the number of opened base pairs at different temperatures. At $\hat T=2.327$, the DNA is unstable against denaturation through
the whole profile (i.e., the curve is monotonously decreasing). State {\it a}, marked by arrow, corresponds to the
intermediate state {\it a} shown in Fig. S1.}
\label{fig:S1}
\end{figure}
\begin{figure}[ht]
\centering
\includegraphics[width=.40\textwidth]{SI3.eps}
\caption{
Occupation probability of L48AS21 plotted against the number of opened
base pairs at different temperatures}
\label{fig:figure2}
\end{figure}
\begin{figure}[h]
        \begin{center}
                \includegraphics[width=.50\textwidth]{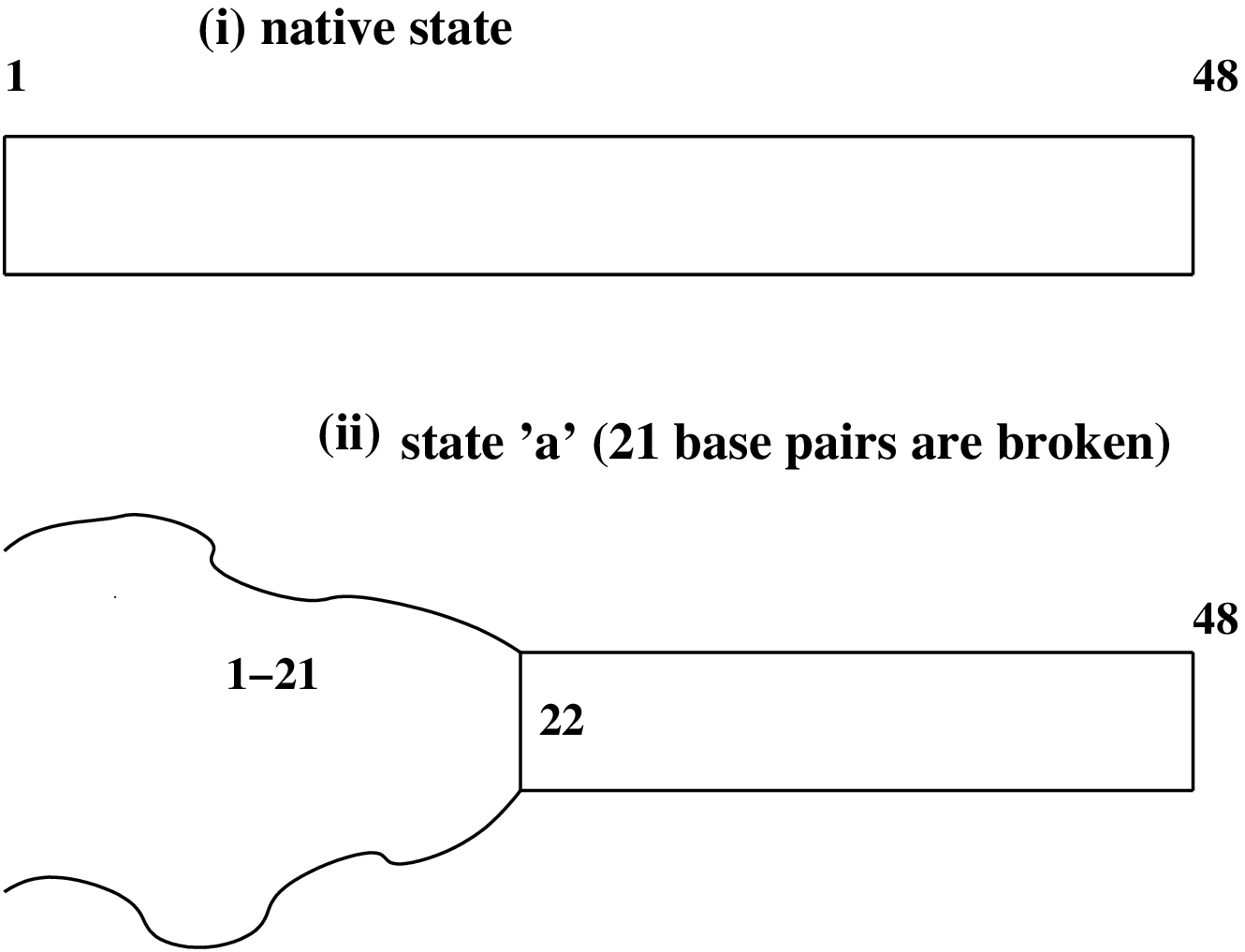}
                \caption{Schematic representation for native and `a' state of L48AS21.}
                \label{fig:example}
        \end{center}
\end{figure}

\begin{figure}[h]
        \begin{center}
                \includegraphics[width=.50\textwidth]{SI5.eps}
                \caption{The solid black line indicates the melting curve for L48AS21. The inset shows the derivative $d<f>/d\hat T$,
with the same temperature scale as in the main figure, to show the two-stage melting more clearly.}
                \label{fig:example}
        \end{center}
\end{figure}

The schematic diagram of (DNA) L48AS21 configuration in the dominant partial
denaturation state {\it a} is shown in Fig. S4, where state a corresponds to broken $21$ bp's, from $1^{st}$ to $21^{th}$ (termed previously as
a bubble at end).

We also show the melting profile in Fig. S5. The inset in Fig. S5 shows the derivative of melting profile.
Note the double shoulder in the plot f($\hat T$) vs. $\hat T$, where the  first shoulder corresponds to denaturation
of first 21 base pairs (from $1^{st}$ to $21^{th}$), whereas the second shoulder corresponds to fully denaturation
(from $22^{th}$ to $48^{th}$). Similar results for the melting curve have been obtained in Ref. [51-53](of the main text) and theoretically verified in Ref. [52] (of the main text), yet no result has been reported for the free-energy.
\section{II. Denaturation of Homosequence DNA}
To ensure our model displays a physically sound behavior we show the denaturation/melting of the homosequence DNA.
\begin{figure}[h]
        \begin{center}
                \includegraphics[width=.45\textwidth]{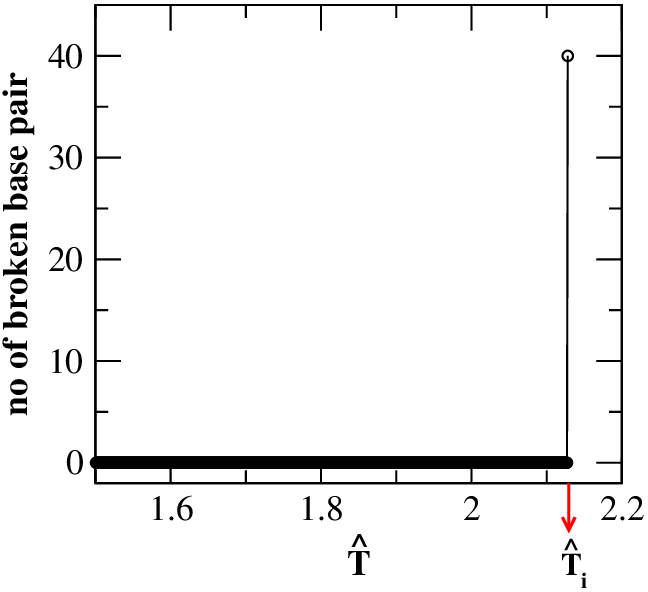}
                \caption{Fig. S6 shows denaturation instability phase diagram for homosequence DNA: Number of opened base pairs
against temperature. We observe all base pair bonds break simultaneously (at $\hat T= \hat T_{i}$) due to an avalanchelike effect.
There is no any intermediate state between native and fully denaturated state.}
                \label{fig:example}
        \end{center}
\end{figure}
Fig. S6 shows the instability phase diagram for homosequence (dGdC) DNA. One can clearly see that the
duplex (dGdC dsDNA) is unstable at $\hat T_i$ $(=2.128)$, where all hydrogen bonds between the two strands simultaneously break. We finds that denaturation starts from both ends, as can be expected by symmetry. In Fig.2(a)(in the main text)  we show the melting profile for dGdC DNA for a range of strand lengths $L$ between 5 and 1000. We can observe that the melting curve sharpens with increasing $L$, as generally expected for phase transitions in finite systems, and similar to the findings of Ref. [49] (in the main text).

Fig. 2(b) (in the main text) shows the derivative of melting profile. The peak in the derivative curve coincides with the melting temperature and narrows with increase of $L$. It is a sign of first order transition (discontinues transition) in the infinite length limit. The variation of melting temperature $\hat T_m$ at different $L$'s is shown in Fig. S7.
\begin{figure}[h]
        \begin{center}
                \includegraphics[width=.40\textwidth]{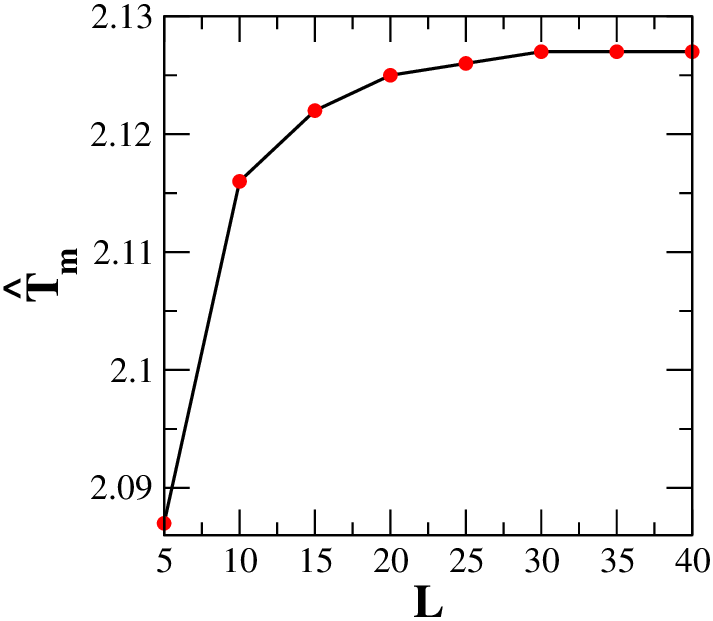}
                \caption{The melting temperature $(\hat T_m)$ $\it vs$ L.}
                \label{fig:example}
        \end{center}
\end{figure}

In Fig. S8(a-f) we show the free-energy profile for homosequence DNA at different temperatures.
Note that no intermediate stable state exists between native and complete denaturation states - until at elevated temperatures
the complete denaturation state becomes the most stable one.

\begin{figure}[t]
        \begin{center}
                \includegraphics[width=.52\textwidth]{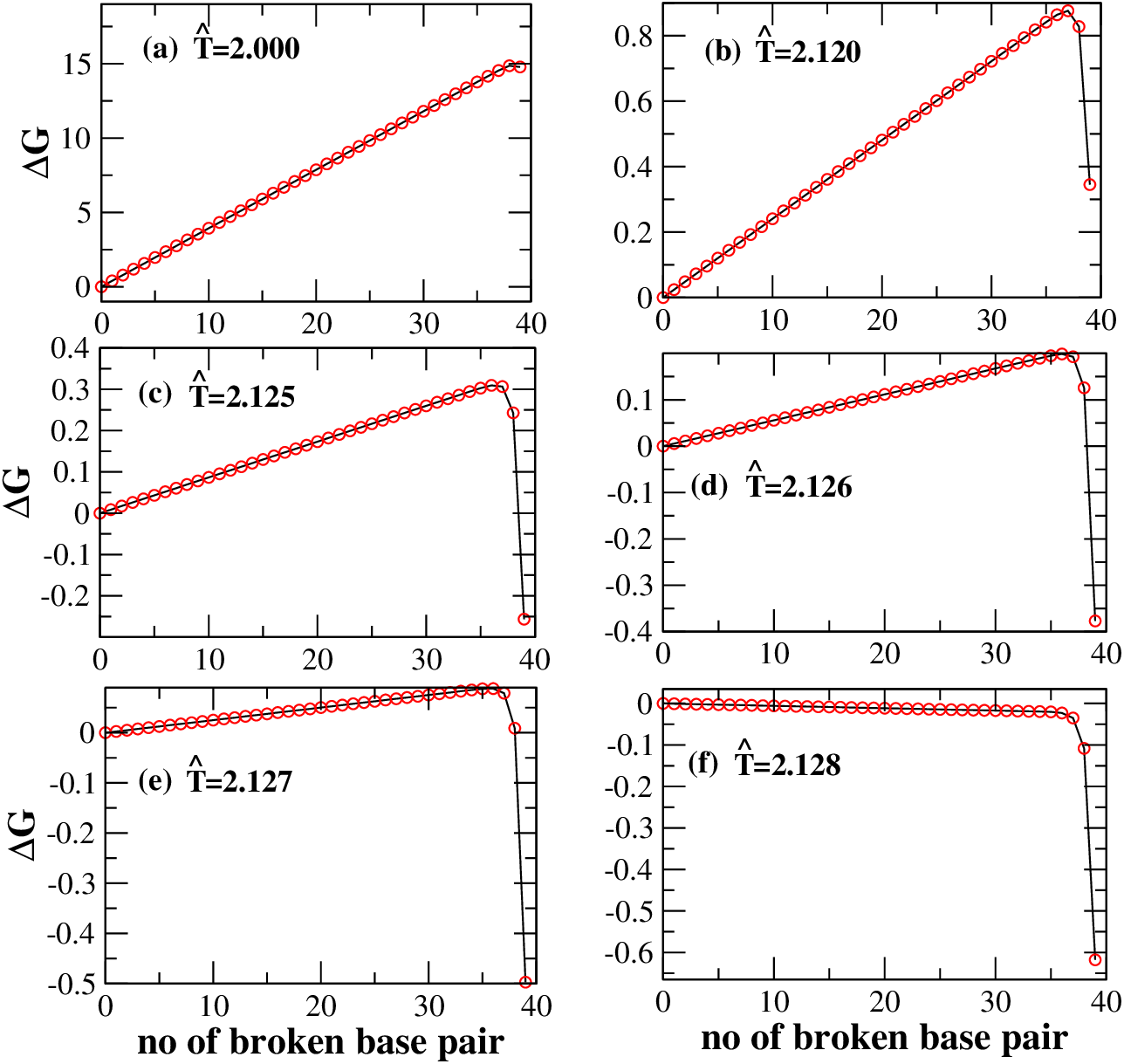}
                \caption{Free-energy landscapes for the  homogeneous dsDNA .}
                \label{fig:example}
        \end{center}
\end{figure}
\newpage
\section{III. Denaturation of diblock DNA}
In this section, we study diblock DNA where half of the strand is made by G-C base-pair whereas the other
half is made by A-T base-pair. We show the melting profile and its derivative for different G-C fractions. Fig. S9(a-k) shows the melting profile (black solid line) and its derivative (red solid line) for the same $L$ but with different G-C fractions, going from 0\% to 100\%. The derivative of melting curve shows two peaks, signifying two transition temperatures, whose relative height changes with increasing G-C fraction, unlike the case of homosequence DNA where only one transition temperature is observed (Fig. S9(a) and Fig. S9(k)). The first peak corresponds to the melting of A-T block and the second peak corresponds to the melting of the G-C block. In Fig. S9(l) we show together the melting profiles for all G-C fractions (blue corresponds to 0\% G-C and brown to 100\%). It is seen that $\hat T_{m}$ increases with increase of G-C fraction, similar to the effect reported in Ref. [50] (in the main text). Moreover, the two transitions become sharper with increase of $L$, see Fig. S10(a-d).

\begin{figure}[t]
        \begin{center}
                \includegraphics[width=.50\textwidth]{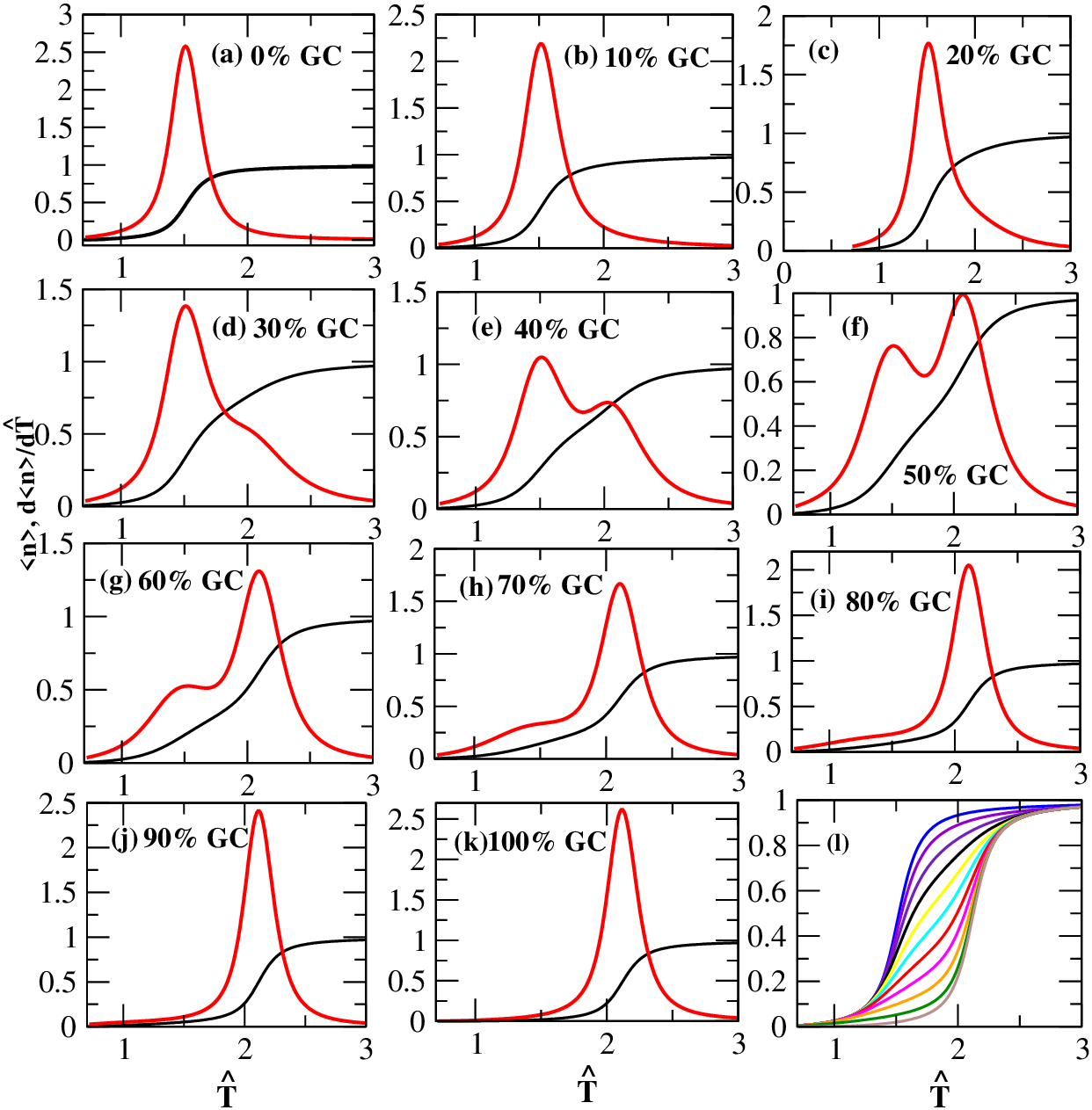}
                \caption{Fig. S9(a-k) shows melting profile (black solid line) and its derivative (black red line) for same L with different G-C concentration (from 0\% - 100\%).}
                \label{fig:example}
        \end{center}
\end{figure}
\begin{figure}[h]
        \begin{center}
                \includegraphics[width=.50\textwidth]{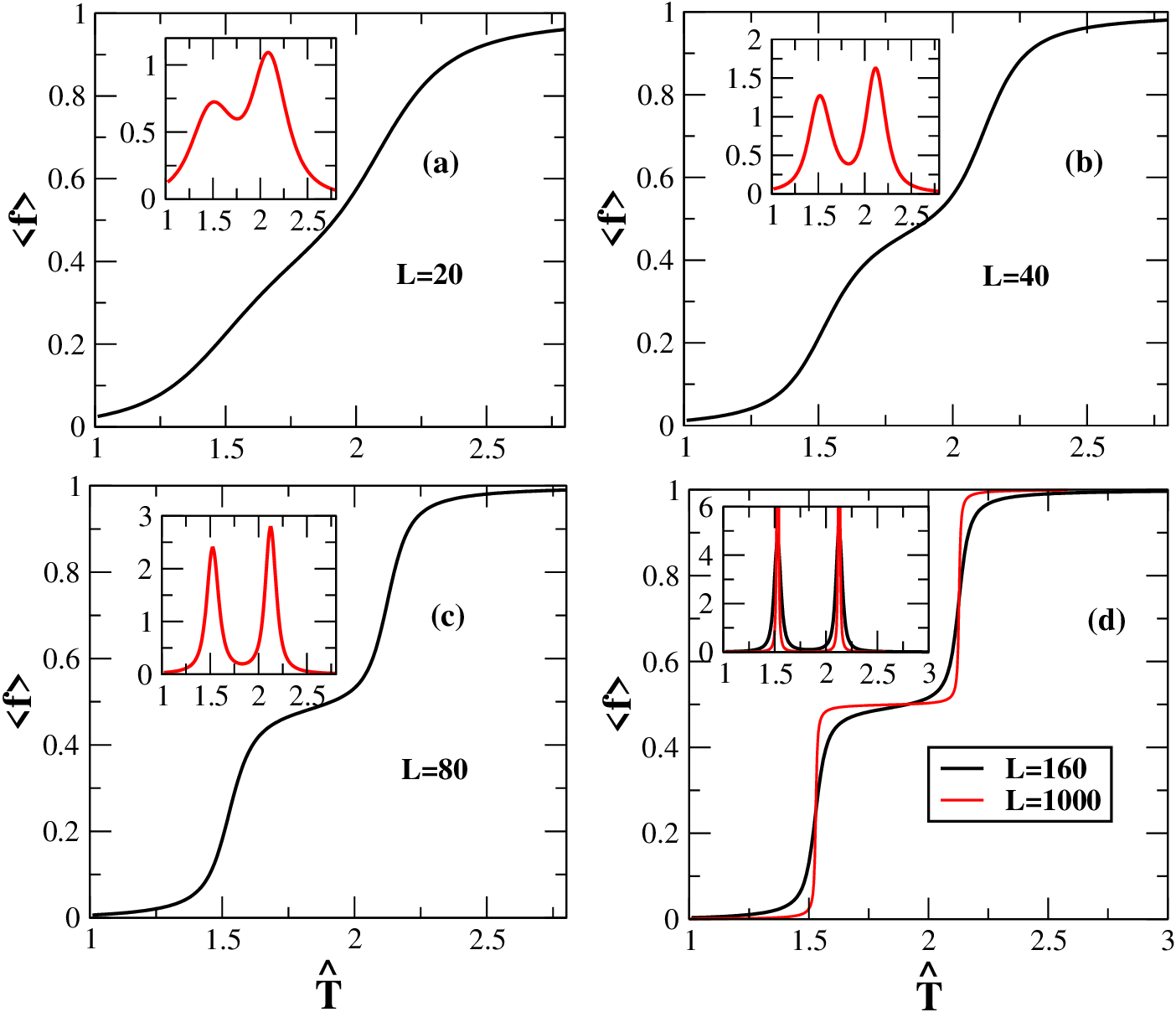}
                \caption{Fig. S10(a-d) shows melting profile and its derivative for diblock DNA (for L=20, 40, 80 160 and 1000).}
                \label{fig:example}
        \end{center}
\end{figure}
\newpage
\section{V.Sequences}
\hspace{-.3cm}{\bf L48AS21:}CATAATACTTTATATTTAATTGGCGGCGCACGGGACCCGTGCGCCGCC\\
{\bf L60B36:}CCGCCAGCGGCGTTATTACATTTAATTCTTAAGTATTATAAGTAATATGGCCGCTGCGCC\\
{\bf L42B18:}CCGCCAGCGGCGTTAATACTTAAGTATTATGGCCGCTGCGCC\\
{\bf L33B9:}CCGCCAGCGGCGTTAATACTTAAGTATTATGGCCGCTGCGCC\\
$\bf (G)_{36} \cdot(C)_{36}:$GGGGGGGGGGGGGGGGGGGGGGGGGGGGGGGGGGGG\\
$\bf (GC)_{18} \cdot (GC)_{18}:$ GCGCGCGCGCGCGCGCGCGCGCGCGCGCGCGCGCGC\\
$\bf (AG)_{18} \cdot (CT)_{18}:$AGAGAGAGAGAGAGAGAGAGAGAGAGAGAGAGAGAG\\
$\bf (AC)_{18} \cdot (GT)_{18}:$ACACACACACACACACACACACACACACACACACAC\\
$\bf (A)_{36} \cdot (T)_{36}:$ AAAAAAAAAAAAAAAAAAAAAAAAAAAAAAAAAAAA\\
$\bf (AT)_{18} \cdot (AT)_{18}:$ATATATATATATATATATATATATATATATATATAT\\
{\bf L234}:CGACGGTATCTCCACCTCACACCTCGAACGGCGGGTCCAACCAACGCAACATCTCAGGCCACACACTAG\\
ACCGCCTTTTTTCTGCCGAGGGCGGCGCACTTGGGGCTCGGCCATTGCTCTAGCCACGCTTTGCGTACGATG\\
AGTCGGAGTACGTGGGAAAGGCCAGGTAGGAACCTGCGAGGAGCGTTGACTGCGTCTGCTTCTTCCCCACTA\\
GTAACTTCTCGCCCTCACTAG\\
{\bf L516}:CCAAGGTTTATACACCTCTAACATCGATAGGCGGATCCATAATAAGCAAATTATAACCCCACACAATT\\
GTCCCACTTTTTTCTGCACTGGCAGCCGCACTTCGGGATAGGAATTTGCTATTCCAACGATTTGAGTTCG\\
ATGACTCGCACTAAGTGGGATTGGCAAGGTTGGAAAATGAGTGCAGCGTTCAATGAGTATGATTATTAAC\\
CACTAGTTTATTATCGCCATAACTTGTAGATCCCTATTCTAGGACCAGTGTGCACAGTAGGTTGATCACA\\
AAAGCGGTGGACTGTATTCACGGTCCCACGAACTATTTTTACGAGACCCTGGTAGTACTCCCCGACTCAG\\
GTTTTCAAATAGGCTGACTAAATCTCCCAACATGTGAAGACAGCATAAAATAACCCAAGACGGCACCAAA\\
TATCGTTCAATTCCGGACAAAGCTCGGCTGGCAACTGATGATTGTGGTAGAGACTGGCAACTAAAAGTTT
TAGGAAGACCTGTAGCACGCTGTCGTAC\\

\section{IV. Melting of L234(59\%G-C) and L516(44\%G-C)}
To compare with experiments as well as with MELTISM programm, we have shown the melting temperature ($T_m$) and the ratio of $T_m$ of
different dsDNA sequences( dsDNA sequences given above in section V) in Table S1 and S2. L234(59\%G-C) and L516(44\%G-C) are two random sequences of 234 $bp's$ and 516 $bp's$, and we drew a single realization of the for each. These realizations are identical in length and G-C concentration to pBR322-234 and pBR322-516 studied in Ref. 5 (in main text), respectively, just that the experimental studies used deterministic sequences. By using the MELTISM software, the melting profile and its derivative for L234 (red line)
and L516 (black line) is shown in Fig. S11(a) and Fig. S11(b), respectively. The ratio of melting temperature from our model, MELTISM, and experiment is given in Table S2 . Comparison with experiment is limited due to the difference between the systems studied.

\begin{figure}[h]
        \begin{center}
                \includegraphics[width=.70\textwidth]{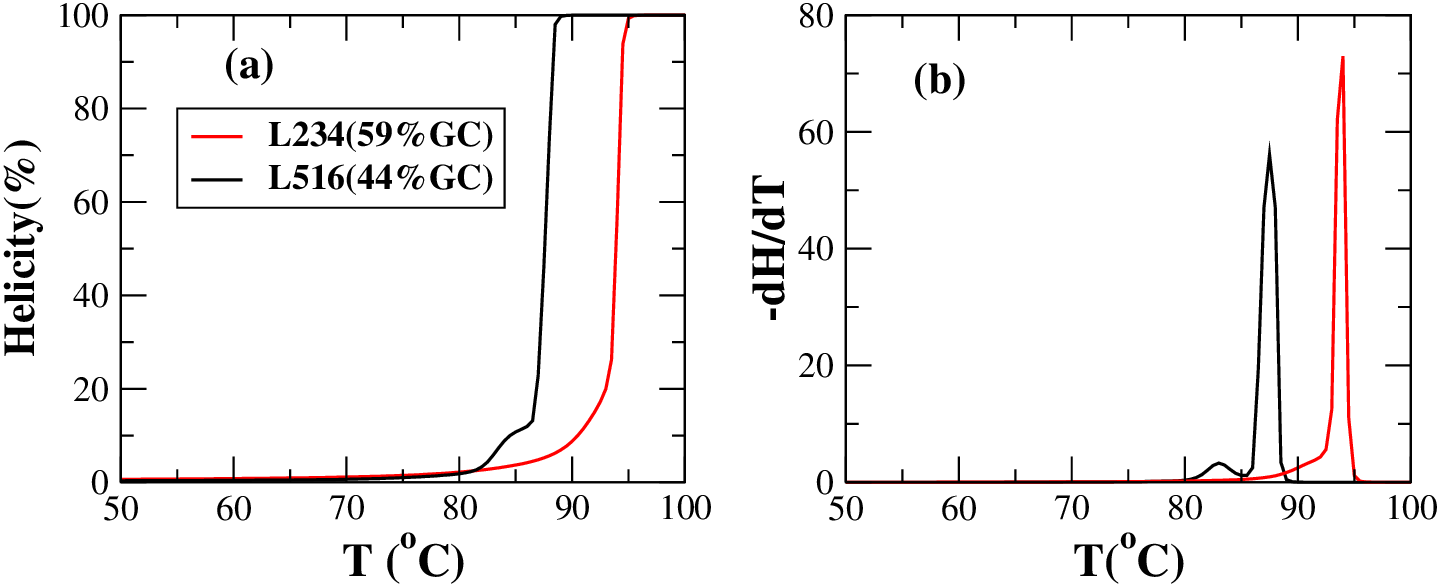}
                \caption{Fig. S11 shows melting profile and its derivative for 59\% GC concentration (L232(red line)) and for
44\% GC concentration (L516(black line)).}
                \label{fig:example}
        \end{center}
\end{figure}

\begin{table}[tbp]
\centering
\caption{Experimentally determined (a: Ref.51, b: Ref.56, c: Ref.5) and calculated melting temperatures  of different dsDNA sequences.}
\label{my-label}
\begin{tabular}{|c|c|c|c|c}
\hline
\begin{tabular}[c]{@{}c@{}}dsDNA\end{tabular} & $T_m$(Experiment) & \begin{tabular}[c]{@{}c@{}} $T_m$(Our model) \end{tabular} & $T_m$(MELTISM) \\
\hline
\begin{tabular}[c]{@{}c@{}} $L33B9$  \end{tabular} & $344K^a$ & \begin{tabular}[c]{@{}c@{}} 2.050 \end{tabular} & 356.6K \\
\hline
\begin{tabular}[c]{@{}c@{}} $L48AS21$ \end{tabular} & $341K^a$ & \begin{tabular}[c]{@{}c@{}} 1.999 \end{tabular} & 361.6K \\
\hline
\begin{tabular}[c]{@{}c@{}} $L42B18$ \end{tabular} & $339K^a$  & \begin{tabular}[c]{@{}c@{}} 1.962 \end{tabular} & 353.9K  \\
\hline
\begin{tabular}[c]{@{}c@{}} $L60B36$  \end{tabular} & $336K^a$  & \begin{tabular}[c]{@{}c@{}} 1.814 \end{tabular} & 351.1K \\
\hline
\begin{tabular}[c]{@{}c@{}} $(GC)_{18}$$\cdot$$(GC)_{18}$ \end{tabular}  & $369K^b$  & \begin{tabular}[c]{@{}c@{}} 2.257  \end{tabular}
& 373.9K \\
\hline
\begin{tabular}[c]{@{}c@{}} $(G)_{36}$$\cdot$$(C)_{36}$ \end{tabular}  & $347K^b$  & \begin{tabular}[c]{@{}c@{}} 2.127 \end{tabular}
& 366.6K \\
\hline
\begin{tabular}[c]{@{}c@{}} $(AT)_{18}$$\cdot$$(AT)_{18}$   \end{tabular}  & $311K^b$  & \begin{tabular}[c]{@{}c@{}} 1.519  \end{tabular}
& 320.1K \\
\hline
\begin{tabular}[c]{@{}c@{}} $(A)_{36}$$\cdot$$(T)_{36}$ \end{tabular}  & $318K^b$  & \begin{tabular}[c]{@{}c@{}} 1.526 \end{tabular}
& 333.2K \\
\hline
\begin{tabular}[c]{@{}c@{}} $(AC)_{18}$$\cdot$$(GT)_{18}$   \end{tabular}  & $340K^b$  & \begin{tabular}[c]{@{}c@{}} 1.936  \end{tabular}
& 352K \\
\hline
\begin{tabular}[c]{@{}c@{}} $(AG)_{18}$$\cdot$$(CT)_{18}$ \end{tabular}  & $332K^b$  & \begin{tabular}[c]{@{}c@{}} 1.925 \end{tabular}
& 347.6K \\
\hline
\begin{tabular}[c]{@{}c@{}} L234(59\%GC) \end{tabular}  & $362.1K^c$  & \begin{tabular}[c]{@{}c@{}} 1.997 \end{tabular}
& 367.7K  \\
\hline
\begin{tabular}[c]{@{}c@{}} L516(44\%GC) \end{tabular}  & $354.6K^c$  & \begin{tabular}[c]{@{}c@{}} 1.895 \end{tabular}
& 358.2K \\
\hline
\end{tabular}
\end{table}

\begin{table}[hbp]
\centering
\caption{The ratio of melting temperature (determined by experiment, Meltism and our model) of different dsDNA sequences.}
\label{my-label}
\begin{tabular}{|c|c|c|c|c}
\hline
\begin{tabular}[c]{@{}c@{}}Ratio of dsDNA\end{tabular} & Ratio of $T_m$(experiment) & \begin{tabular}[c]{@{}c@{}} Ratio of $T_m$(our model) \end{tabular} & Ratio of $T_m$(Meltism) \\
\hline
\begin{tabular}[c]{@{}c@{}} $(L48AS21)_{Tm}$/$(L60B36)_{Tm}$  \end{tabular} & 1.01 & \begin{tabular}[c]{@{}c@{}} 1.10 \end{tabular} & 1.03 \\
\hline
\begin{tabular}[c]{@{}c@{}} $(L33B9)_{Tm}$/$(L60B36)_{Tm}$ \end{tabular} & 1.02 & \begin{tabular}[c]{@{}c@{}} 1.13 \end{tabular} & 1.02 \\
\hline
\begin{tabular}[c]{@{}c@{}} $(L42B18)_{Tm}$/$(L60B36)_{Tm}$ \end{tabular} & 1.01  & \begin{tabular}[c]{@{}c@{}} 1.08 \end{tabular} & 1.01 \\
\hline
\begin{tabular}[c]{@{}c@{}} $(L48AS21)_{Tm}$/$(L42B18)_{Tm}$   \end{tabular} & 1.01  & \begin{tabular}[c]{@{}c@{}} 1.02 \end{tabular} & 1.02 \\
\hline
\begin{tabular}[c]{@{}c@{}} $(L33B9)_{Tm}$/$(L42B9)_{Tm}$  \end{tabular}  & 1.01  & \begin{tabular}[c]{@{}c@{}} 1.04  \end{tabular}
& 1.01 \\
\hline
\begin{tabular}[c]{@{}c@{}} $(L33B9)_{Tm}$/$(L48AS21)_{Tm}$ \end{tabular}  & 1.01  & \begin{tabular}[c]{@{}c@{}} 1.02 \end{tabular}
& .99 \\
\hline
\begin{tabular}[c]{@{}c@{}} $(GC)_{18}$$\cdot$$(GC)_{18}$/$(G)_{36}$ $\cdot$$(C)_{36}$   \end{tabular}  & 1.06  & \begin{tabular}[c]{@{}c@{}} 1.06  \end{tabular} & 1.02 \\
\hline
\begin{tabular}[c]{@{}c@{}} $(A)_{36}$$\cdot$$(T)_{36}$/$(AT)_{18}$$\cdot$$(AT)_{18}$  \end{tabular}  & 1.02  & \begin{tabular}[c]{@{}c@{}} 1.01 \end{tabular} & 1.04 \\
\hline
\begin{tabular}[c]{@{}c@{}}  $(AC)_{18}$$\cdot$$(GT)_{18}$/$(AG)_{18}$$\cdot$$(CT)_{18}$   \end{tabular}  & 1.02  & \begin{tabular}[c]{@{}c@{}} 1.01  \end{tabular} & 1.01 \\
\hline
\begin{tabular}[c]{@{}c@{}} $(L234)_{Tm}$/$(L516)_{Tm}$ \end{tabular}  & 1.02  & \begin{tabular}[c]{@{}c@{}} 1.05 \end{tabular}
& 1.03 \\
\hline
\end{tabular}
\end{table}

\begin{table}[tbp]
\centering
\caption{Spring constant for covalent bond(m), base pair(a , b ) and sixteen possible DNA base-pair stacking $(d_{\nu})$patterns.}
\label{my-label}
\begin{tabular}{|c|c|c|c|c|c}
\hline
\begin{tabular}[c]{@{}c@{}}m,a,b,$d_{\nu}$\end{tabular} & Spring constant & \begin{tabular}[c]{@{}c@{}}m,a,b,$d_{\nu}$ \end{tabular} & Spring constant \\
\hline
\begin{tabular}[c]{@{}c@{}} covalent bond \end{tabular} & $m= 10$  & \begin{tabular}[c]{@{}c@{}}  \end{tabular} &  \\
\hline
\begin{tabular}[c]{@{}c@{}} A-T(base-pair) \end{tabular} & $a= 1.83$  & \begin{tabular}[c]{@{}c@{}} G-C (base -pair) \end{tabular} & b=4.39 \\
\hline
\begin{tabular}[c]{@{}c@{}}  C$\cdot$G \\ G$\cdot$C \end{tabular}               & $d_1$ = 1.905                              & \begin{tabular}[c]{@{}c@{}} C$\cdot$G \hspace{.3cm}  T$\cdot$A\\  A$\cdot$T \hspace{.3cm}  G$\cdot$C\end{tabular} & $d_2$,
$d_3$ = 1.375 \\
\hline
\begin{tabular}[c]{@{}c@{}}C$\cdot$G \hspace{.3cm}   A$\cdot$T\\ T$\cdot$A \hspace{.3cm}   C$\cdot$G\end{tabular}      & $d_{4}, d_{5}$ = 1.28                              & \begin{tabular}[c]{@{}c@{}}G$\cdot$C\\ C$\cdot$G\end{tabular}                  & $d_6$ = 1.265                              \\
\hline
\begin{tabular}[c]{@{}c@{}}G$\cdot$C \hspace{.3cm}  C$\cdot$G\\ G$\cdot$C \hspace{.3cm}    C$\cdot$G\end{tabular}    & $d_{7},d_{8} $  1.08                             & \begin{tabular}[c]{@{}c@{}}T$\cdot$A\\ A$\cdot$T\end{tabular}                  & $d_9 $ = 0.86                              \\
\hline
\begin{tabular}[c]{@{}c@{}}G$\cdot$C \hspace{.3cm}   A$\cdot$T\\ T$\cdot$A \hspace{.3cm}   C$\cdot$G\end{tabular} & $d_{10}, d_{11} $ = 0.86                              & \begin{tabular}[c]{@{}c@{}}G$\cdot$C \hspace{.3cm}    T$\cdot$A\\ A$\cdot$T \hspace{.3cm}   C$\cdot$G\end{tabular}     & $d_{12}, d_{13} $ = 0.89                              \\
\hline
\begin{tabular}[c]{@{}c@{}}A$\cdot$T \hspace{.3cm}   T$\cdot$A\\ A$\cdot$T \hspace{.3cm} T$\cdot$A\end{tabular}  & $d_{14}, d_{15} $ = 0.705                               & \begin{tabular}[c]{@{}c@{}}A$\cdot$T\\ T$\cdot$A\end{tabular}                  & $d_{16} $ = 0.5                                \\
\hline
\end{tabular}
\end{table}